\documentclass[a4paper]{article}

\usepackage{INTERSPEECH2022}

\usepackage{diagbox}
\usepackage{multirow}
\usepackage{amsmath}
\usepackage{xcolor}

\title{QbyE-MLPMixer: Query-by-Example Open-Vocabulary Keyword Spotting using MLPMixer}

\name{Jinmiao Huang$^{1*}$\thanks{*Authors contributed equally}, Waseem Gharbieh$^{1*}$, Qianhui Wan$^{1*}$, Han Suk Shim$^2$, Chul Lee$^2$}
\address{
  $^1$LG Electronics Toronto AI Lab\\
  $^2$LG Electronics Artificial Intelligence Lab}
\email{jinmiao.huang@lge.com, waseem.gharbieh@lge.com, qianhui.wan@lge.com, hansuk.shim@lge.com, 
clee.lee@lge.com}

\begin{document}

\maketitle
\begin{abstract}
Current keyword spotting systems are typically trained with a large amount of pre-defined keywords. Recognizing keywords in an open-vocabulary setting is essential for personalizing smart device interaction. Towards this goal, we propose a pure MLP-based neural network that is based on MLPMixer - an MLP model architecture that effectively replaces the attention mechanism in Vision Transformers. We investigate different ways of adapting the MLPMixer architecture to the QbyE open-vocabulary keyword spotting task. Comparisons with the state-of-the-art RNN and CNN models show that our method achieves better performance in challenging situations (10dB and 6dB environments) on both the publicly available Hey-Snips dataset and a larger scale internal dataset with 400 speakers. Our proposed model also has a smaller number of parameters and MACs compared to the baseline models.

\end{abstract}


\section{Introduction}
Traditional keyword spotting relies on fixed wake words such as ``Hey Siri", ``Alexa", or ``Ok Google". By allowing a user to set their own custom wake word, we enable greater degree of flexibility and personalization. This is the essence of user-defined keyword spotting but it comes with many challenges including low latency, small memory footprint, and utterances that are out of the training distribution. 

Over the years, different neural networks have been proposed for the fixed keyword spotting (KWS) task: for example, using Deep Neural Networks (DNNs) \cite{chen2014small, prabhavalkar2015automatic}, Time Delay Neural Networks \cite{bai2019time}, Convolutional Neural Networks (CNNs) \cite{sainath2015convolutional}, Recurrent Neural Networks (RNNs) \cite{fernandez2007application, yamamoto2019small} or transformers \cite{adya2020hybrid, wang2021wake}. However, a large amount of target keyword data is required to effectively train those models. On the other hand, for open-vocabulary keyword spotting, early work relied on the output of an Automated Speech Recognition (ASR) system. For example, by looking for transcript matching on the word \cite{miller2007rapid} or phoneme \cite{brown1997open} level. These systems are computationally expensive and usually show performance degradation when the keywords are out of the vocabulary. Recent work in open-vocabulary keyword spotting focuses on mapping variable-duration audio signals to a fixed-length embedding in vector space. This approach is known as Query By Example (QbyE). In a QbyE system, RNNs are generally used as an encoder to extract the keyword embedding, for example, \cite{chen2015query} is the first to use Long Short Term Memory (LSTM) networks \cite{hochreiter1997long} for QbyE open-vocabulary keyword spotting problem. \cite{lugosch2018donut, bluche2020small, zhuang2016unrestricted} use RNNs with Connectionist Temporal Classification (CTC) loss to predict keyword sequences.  \cite{huang2021query} uses Gated Recurrent Unit (GRU) with a multi-head self-attention mechanism and softtriple loss.   
Non QbyE efforts on this task include the use of a Siamese network with triplet hinge loss \cite{settle2016discriminative}, Recurrent Neural Network Transducer (RNN-T) model to predict phoneme or grapheme based subword units \cite{he2017streaming}, and a metric-based meta-learning algorithm called Prototypical Networks with Max-Mahalanobis Center loss to tackle the scenario where users can define various number of spoken terms \cite{chen21u_interspeech}.  

The MLPMixer \cite{tolstikhin2021mlp} is a recently proposed alternative to the Vision Transformer (ViT) \cite{dosovitskiy2020image}. It uses Multi Layer Perceptrons (MLPs) exclusively to replace the attention mechanism in ViT. Other similar MLP-based models also achieve competitive performance on vision \cite{liu2021pay, touvron2021resmlp, ding2021repmlp} and language \cite{liu2021pay} tasks. That being said, we find the MLPMixer's simple and intuitive design to be appealing so we decided to adapt it to the QbyE open-vocabulary KWS task.

We compare the MLPMixer's performance to an RNN baseline model and some popular CNN models. Our results show that the MLPMixer model is able to outperform those models using a smaller number of parameters and multiply-accumulates (MACs), which indicates that the MLPMixer can also serve as a strong alternative model for audio applications. Our contributions are the following: (1) We propose a simple yet effective adaptation for the MLPMixer for the QbyE KWS problem. (2) Our adaptation to audio provides a significant performance boost compared to the original MLPMixer model. (3) We show that the MLPMixer outperforms other state-of-the-art CNN and RNN based models. (4) We used the publicly available Hey-Snips dataset to conduct experiments on both non far-field and far-field environments, which can serve as a foundation for future work on open-vocabulary keyword spotting.  



\section{Methods}
\begin{figure*}[htb]
  \centering
  \includegraphics[width=0.9\textwidth]{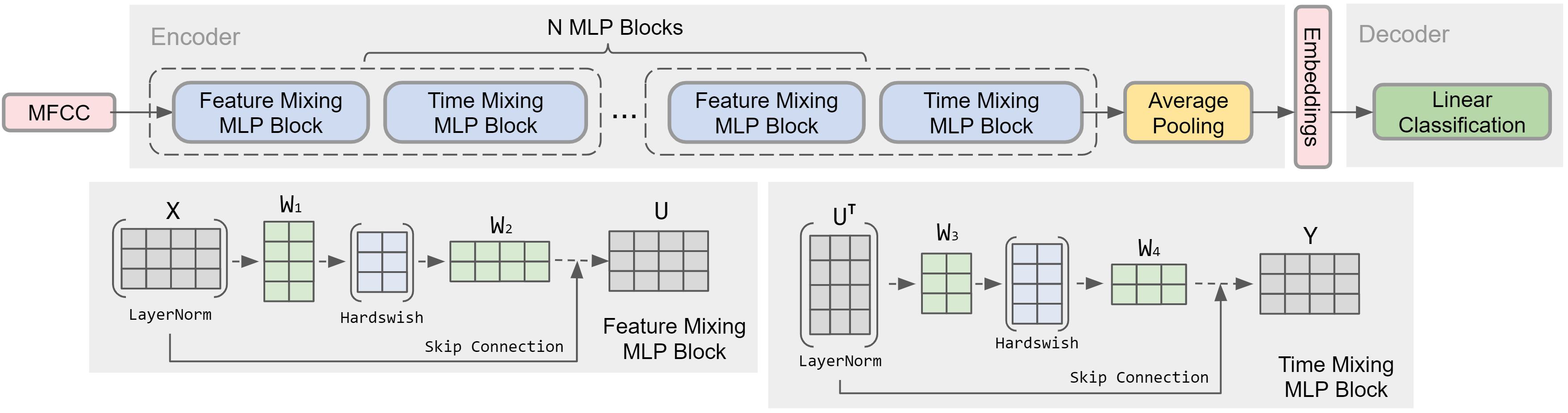}
\caption{Encoder-Decoder System Architecture with MLP Blocks}
\label{fig:system}
\end{figure*}

The encoder-decoder structure can be naturally applied to a QbyE system. The encoder compress data from high dimension into low dimensional embeddings. The decoder is used to tie the encoder to loss functions so embeddings belonging to the same class will be closer to each other while embeddings belonging to different classes will be further apart. During inference, the decoder is completely dropped and the triggering decision is made by comparing the distances between the enrolled and query embeddings. Figure \ref{fig:system} shows the system architecture.


\subsection{Input Representation}
We used Mel-frequency Cepstral Coefficients (MFCCs) as the input features. We extract 81-dimensional MFCC features from 1s long audio computed every 12.5 ms with a window of 25 ms. This results in a matrix with 81 dimensions in both feature and temporal space. CMVN is then applied on the MFCC's temporal dimension before being input to the encoder. 

Since the ViT \cite{dosovitskiy2020image}, many models including the MLPMixer split the input into non-overlapping patches and compute the embedding of each patch before feeding it to the model. In this paper, we keep the feature and timestep dimensions intact and use the 81 features across 81 timesteps directly as input to our model. There are two reasons behind this: firstly, we wanted to study the impact of input representation on the model's ability to extract information from the entire feature space. Secondly, providing the input directly reduces the model's computational and memory footprint. Our experiments show that this approach is more effective than patching (See details in \ref{sec:ablation_input}). 

\subsection{MLPMixer}
\label{sec:mlpmixer}

Unlike the original MLPMixer which first operates on flattened image embedding patches (token-mixing) and then on the channel dimension (channel-mixing), we apply the mixing on the feature and temporal dimensions. Specifically, we directly apply a linear layer to the normalized features to project the feature dimension $f$ to a hidden dimension $h$, followed by another linear layer to project the hidden dimension $h$ back to $f$. The two linear layers form the major components of an MLP block. We call the first MLP block feature-mixing. Similarly, we apply two linear layers on the temporal space to project the time dimension $t$ to a hidden dimension $g$ and then project $g$ back to $t$, we call this second MLP block time-mixing. For a given input $\mathbf{X}\in \mathbb{R}^{f\times t}$ (since there is only one channel, the channel dimension is omitted to simplify the notation), the feature mixing MLP block is implemented as follows:
\begin{equation}
\label{eq:mixer}
    \mathbf{U}= \mathbf{X} +  \mathbf{W}_{2}\sigma(\mathbf{W}_{1}\text{LayerNorm}(\mathbf{X})) \\
\end{equation}
where $\mathbf{X}\in\mathbb{R}^{f\times t}$ denotes the extracted features, $\mathbf{W}_1\in\mathbb{R}^{h\times f}$ and $\mathbf{W}_2\in\mathbb{R}^{f\times h}$ are the weights in the first and second linear layer in an MLP block respectively. We also added LayerNorm and activation function $\sigma$ in each MLP block. We use Hardswish \cite{howard2019searching} as the activation function in this paper as it shows better performance than other activation functions (see detailed comparisons in section \ref{sec:ablation_activation}). A residual connection is added to link the projection from MLP block with the original input $\mathbf{X}$ to form the final projection $\mathbf{U}\in \mathbb{R}^{f\times t}$ on the feature dimension. We then use $\mathbf{U}$ as the input to the time mixing MLP block. Similar operations were added to $\mathbf{U}^T$ to extract information on the temporal dimension:
\begin{equation}
\label{eq:mixer2}
    \mathbf{Y}=\mathbf{U} + \left(\mathbf{W}_{4}\sigma(\mathbf{W}_{3}\text{LayerNorm}(\mathbf{U}^T))\right)^T \\
\end{equation}
where $\mathbf{W}_3\in\mathbb{R}^{g\times t}$ and $\mathbf{W}_4\in\mathbb{R}^{t\times g}$ are the weights of the two fully-connected layers in the time-mixing MLP block. $\mathbf{Y}\in\mathbb{R}^{f\times t}$ is the projection on the temporal dimension. 

Each MLP block generates the same output size as its input. For example, regardless of how we setup the dimension of the hidden space $h$ and $g$ in each MLP block, the output from MLP block is always $f \times t$. Due to this design, MLP blocks can be easily stacked together and we only need to tune parameters $h$, $g$, number of stacked blocks $n$, and the activation function $\sigma$, which makes hyperparameter optimization straightforward.

The output of the MLPMixer is passed to an average pooling layer to aggregate the information on the temporal dimension. Formally, for an output $\mathbf{O} \in \mathbb{R}^{f\times t}$ from the last MLP block, we apply the average pooling operation on the temporal dimension to generate the embedding $z\in\mathbb{R}^{f}$: 
\begin{equation}
z=\frac{1}{t} \sum_{i=1}^{t} \mathbf{O}_{:, i}
\end{equation}
%


\subsection{Inference}
During inference, the linear layer from the decoder is dropped and the embedding output from the average pooling layer is used. Since the model is trained on 1s long audio samples, for any input audio, we take a 1s moving window with a stride of 100 ms and feed it to the network. The cosine distance is then used to compare the similarity between the embedding vectors generated from query audio and each of the $n$ embedding vectors saved from enrollment. When calculating the cosine distance, if the enrollment embedding is shorter than the query embedding, we convolve the enrollment embedding with the query embedding and take the minimum distance. Otherwise, if the enrollment embedding is longer than the query embedding, we zero pad the left side of the query embedding vector to match the size of the enrollment embedding. Note that in a streaming setting, padding is not necessary, since we can create a buffer with the same size as the longest enrollment so the query always matches the length of the longest enrollment. After computing the cosine distance for all enrollments, the minimum value is taken and compared to a threshold to make a triggering decision. If the value is smaller than the set threshold, the system triggers a positive response. For our system, the embedding size is the same as the MFCC feature dimensions (81), which is considerably small compared to other baseline models which have an embedding size on the order of 1000 (e.g. MobilenetV3: 960). The small embedding size also shows that the MLPMixer is able to effectively project the useful information to a small hidden space.

\section{Experiments}


\subsection{Training}
We used the same protocol as \cite{huang2021query} to train our model. We used the Librispeech \cite{panayotov2015librispeech} dataset which is comprised of 1000 hours of read English audiobooks sampled at 16 kHz along with annotated text. Since we want to train our model to classify the word given the audio, we use the Montreal Forced Aligner \cite{mcauliffe2017montreal} tool to generate word level annotated segmentations. After segmenting the audio, we standardize it to be 1s long by random clipping or zero padding on both sides of the audio depending on whether the audio is longer or shorter than 1s.


We augmented the training data with random 4 to 12 dB background noise from the ``noise-train" folder in the Microsoft Scalable Noisy Speech Dataset (MS-SNSD) \cite{reddy2019scalable}. In addition, we simulated effects of far-field conditions following \cite{ko2017study} with 50\% probability of adding 4dB to 15dB point source noise on top of the far-field effect. Cross entropy loss was used as the objective function.

\subsection{Evaluation}

For the positive queries, we used the test portion of the Hey-Snips dataset and an internal dataset to test our model. The test portion of the Hey-Snips dataset contains 2,588 positive utterances from 520 speakers with a maximum of 10 utterances per speaker. Since we need multiple keyword recordings from the same speaker to perform the evaluation, we selected all the speakers with 10 keyword utterances and used three random utterances for enrollment and the remaining seven utterances for query. This reduced the number of speakers down to 40. The internal dataset contains 8 keywords named after actual home appliances: ``LG Styler", ``LG Washer", ``LG Dryer", ``Hey LG", ``LG Fridge", ``LG Puricare", ``LG Oven", and ``Hey Cloi". Each keyword was uttered 10 times by 50 speakers. Similar to Hey-Snips, we used three random utterances for enrollment and the remaining seven for query. Both datasets were recorded in a clean environment. We synthetically distorted both datasets in order to simulate real-world environment by adding noise and far-field effects to them. Specifically, we used the ``noise-test" folder from the MS-SNSD dataset to add 10dB and 6dB noises. We also added far-field effects to the clean dataset before adding 10dB and 6dB noise to it. This resulted in six scenarios in total.     

For the negative queries, we used the negative samples from Hey-Snips test set. It contains about 20K utterances of general sentences from 1,469 speakers with a maximum of 30 utterances per speaker. We split the negative set into 2 halves and randomly add 10dB and 6dB noise to each half. The ``noise-test" folder in the MS-SNSD dataset was used for adding noise. In the far-field case, we added far-field effects to the negative set before adding noise. Each set was evaluated alongside its corresponding positive queries.


\section{Results}

\subsection{Baseline Models}

For baseline models, we first choose a state-of-the-art RNN model with GRU and self-attention (GRU-ATTN) \cite{huang2021query} which previously reported the open-vocabulary keyword spotting results on the Hey-Snips dataset. 

In addition, since computer vision models show good performance on many audio tasks (for example, MobileNetV2 \cite{sandler2018mobilenetv2} were used for audio tagging \cite{kong2020panns}, pre-defined keyword spotting \cite{tang2020howl} and personalized keyword spotting \cite{jia20212020}), we use them as baseline models to compare our model against. Specifically, we experimented with the edge friendly MobileNetV2 \cite{sandler2018mobilenetv2}, MobileNetV3 \cite{howard2019searching}, and EfficientNetB0 \cite{tan2019efficientnet}. We also included ViT in our baseline, as it shows competitive performance compared to state-of-the-art CNN models in vision tasks. 

The inputs to vision models are usually images with three color channels, whereas our input has only one channel. In order to adapt the vision models to our task, we adopt the same preprocessing technique from \cite{tang2020howl} to map one-channel audio into three channels. We found that models with ImageNet \cite{deng2009imagenet} pretrained weights show significantly better performance than the ones without, even though there is no clear relationship between MFCC features and ImageNet samples. Similar observations on the effectiveness of transfer learning from ImageNet pretrained models on audio tasks is also reported in \cite{palanisamy2020rethinking}.

\subsection{Results and Size Comparison}

When designing our MLPMixer model, our objective was to answer the following question: Can we find a model architecture that has competitive performance while keeping the number of parameters and MACs inline with the baseline models? Table \ref{tab:model_size} shows that among the baseline models, GRU-ATTN \cite{huang2021query} has the smallest number of parameters (550K) while MobileNetV3 \cite{howard2019searching} has the smallest number of MACs (22.24M). Under this criterion, we ran hyperparameter optimization trials using Asynchronous Hyperband Search (ASHA) with the intention of finding a model that has a smaller footprint than the baseline models. Our hyperparameter search with hundreds of sampled experiments yielded a model with 250K parameters and 20M MACs. This model is constructed with 12 mixing blocks containing 64 hidden layers for both the time and frequency mixing blocks and no dropout. Our implementation of the MLPMixer is able to outperform MobileNetV3 in most cases, especially under challenging conditions.

\begin{table}[tbh]
\centering
\caption{Number of parameters and MACs for baseline models and our MLPMixer model. For baseline models, MobileNet family has a lower number of MACs while the parameter size is significantly larger than GRU-ATTN. Our model has the lowest number of parameters and MACs among all models.}
\resizebox{0.7\columnwidth}{!}{
\label{tab:model_size}
\begin{tabular}{ccc}
Model            & Params (M) & MACs (M) \\ \hline
GRU-ATTN  \cite{huang2021query}           & 0.55       & 41.23   \\
MobileNetV2 \cite{sandler2018mobilenetv2}    & 2.22      & 29.22   \\
MobileNetV3 \cite{howard2019searching}    & 2.97      & 22.24   \\
EfficientnetB0 \cite{tan2019efficientnet}      & 4.01 & 39.06 \\
ViT \cite{dosovitskiy2020image} & 0.96        & 77.06      \\
QbyE-MLPMixer         & \textbf{0.25}         & \textbf{20.16}       \\ \hline
\end{tabular}
}
\end{table}

Tables \ref{tab:hey_snips} and \ref{tab:hey_snapdragon} show the False Rejection Rate (FRR) at 0.3 False Acceptance (FA) per hour for Hey-Snips and the internal dataset respectively. In general, our model gives lower FRRs compared to the baseline models under more challenging conditions. More specifically, compared to the best performance among baseline models on Hey-Snips dataset, our model shows 2.15\% and 4.29\% decrease in FRR under on non far-field 10dB and 6dB condition, and 6.43\% and 8.57\% decrease under far-field 10dB and 6dB condition. For the internal dataset, MobileNetV3 shows the best performance when there is no additional noise added, whereas our model gives better FRRs under more challenging conditions.

\begin{table}[tbh]
\captionof{table}{FRR (\%) at 0.3 FAs per hour for clean, 10dB and 6dB on Hey-Snips dataset}
\resizebox{\columnwidth}{!}{

\begin{tabular}{cccc|ccc}

\multirow{2}{*}{Model} & \multicolumn{3}{c|}{Non Far-Field}                                      & \multicolumn{3}{c}{Far-Field}                                                     \\ \cline{2-7} 
                       & clean                    & 10dB                      & 6dB                       & clean                     & 10dB                      & 6dB                       \\ \hline
GRU-ATTN \cite{huang2021query}                & 7.86 & 11.79 & 15.71 & 20.00 & 20.64 & 32.50 \\
MobileNetV2 \cite{sandler2018mobilenetv2}          & \textbf{5.36} & 12.50 & 20.71 & 14.64 & 15.00 & 35.36\\
MobileNetV3 \cite{howard2019searching}          & 5.71 & 13.93 & 21.79 & 12.86 & 18.21 & 38.93 \\
EfficientnetB0 \cite{tan2019efficientnet}      & 5.71 & 12.86 & 18.21 & 16.43 & 13.93 & 36.79 \\
ViT \cite{dosovitskiy2020image} & 8.21 & 10.36 & 14.29 & 12.14 & 15.36 & 28.93\\
QbyE-MLPMixer   & \textbf{5.36} & \textbf{8.21} & \textbf{10.00} & \textbf{8.93}  & \textbf{7.50} & \textbf{20.36}  \\ \hline
\end{tabular}
}
\label{tab:hey_snips}
\end{table}

\begin{table}[tbh]
\captionof{table}{FRR (\%) at 0.3 FAs per hour for clean, 10dB and 6dB on internal dataset}
\resizebox{\columnwidth}{!}{
\begin{tabular}{cccc|ccc}
\multirow{2}{*}{Model} & \multicolumn{3}{c|}{Non Far-Field}                                      & \multicolumn{3}{c}{Far-Field}                                                     \\ \cline{2-7} 
                       & clean                    & 10dB                      & 6dB                       & clean                     & 10dB                      & 6dB                       \\ \hline
GRU-ATTN \cite{huang2021query}               & 0.81 & 4.43 & 7.06 & 8.01 & 18.63 & 26.78 \\
MobileNetV2 \cite{sandler2018mobilenetv2}          & 3.54 & 5.38 & 6.84 & 11.25 & 15.44 & 20.09 \\
MobileNetV3 \cite{howard2019searching}          & \textbf{0.62} & \textbf{3.69} & 5.82 & \textbf{3.85} & 11.35 & 16.82 \\
EfficientnetB0 \cite{tan2019efficientnet}      & 2.12 & 4.60 & 6.43 & 5.71 & 11.76 & 16.13 \\
ViT \cite{dosovitskiy2020image} & 4.59 & 5.06 & 6.10 & 9.03 & 13.75 & 16.13 \\
QbyE-MLPMixer  & 2.68 & 3.71 & \textbf{4.32} & 5.29 & \textbf{9.25} & \textbf{11.34} \\ \hline
\end{tabular}
}
\label{tab:hey_snapdragon}
\end{table}

\begin{figure}[htb]
  \centering
  \includegraphics[width=8.2cm]{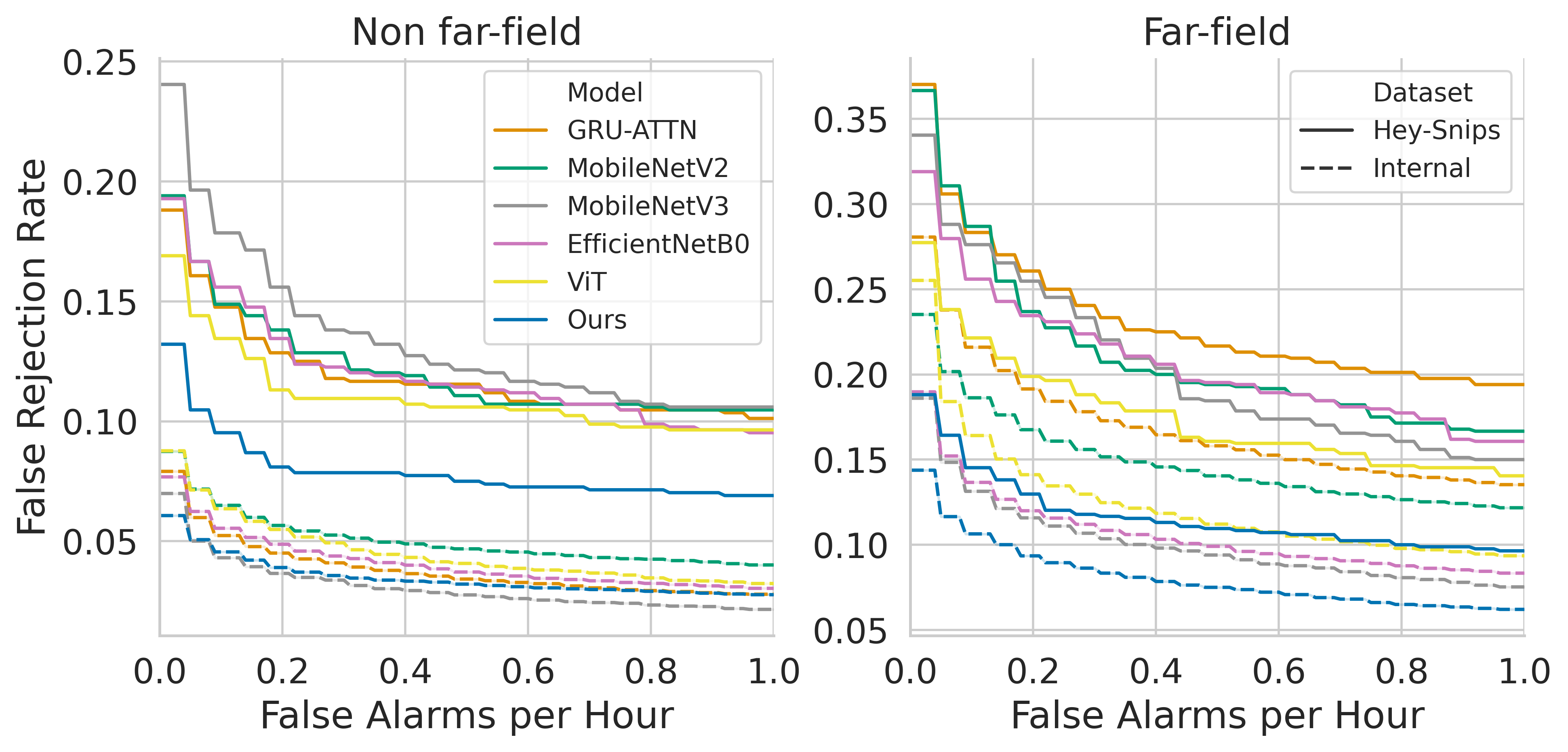}
\caption{ROC on Hey-Snips and Internal datasets}
\label{fig:ROC}
\end{figure}

\subsection{Ablation Study}

\subsubsection{Alternative Input Representations}
\label{sec:ablation_input}

The original MLPMixer model divides the input image into non-overlapping patches and then computes a linear embedding for each patch. Although this procedure is effective for images, we observed that it is suboptimal for audio. Instead, we show that feeding the MFCC features directly to the model to be much more effective. To mimic the original MLPMixer, we divide the MFCC features into 9$\times $9 non-overlapping patches and compute an 81 dimensional linear embedding for each patch (``with PE" in Table \ref{tab:input_rep}). This way the input shape is identical to the input to our proposed model. The patch embedding layer increases the model's number of parameters by 7K and MACs by 500K. For completeness, we also experimented with dividing the MFCC features into 9$\times 9$ non-overlapping patches and reshaping them (without computing linear embeddings) to match the input shape of our proposed model (``w/o PE" in Table \ref{tab:input_rep}). 

Table \ref{tab:input_rep} shows the results of our experiments. We ran the models on all six scenarios (clean, 10dB, and 6dB noise on the non far-field and far-field conditions) on the Hey-Snips and Internal datasets and report the mean FRR for the non far-field and far-field conditions separately.

\begin{table}[tbh]
\captionof{table}{Mean FRR (\%) at 0.3 FAs per hour for various input representations on the Hey-Snips and Internal datasets. PE stands for patch embeddings.}
\label{tab:input_rep}
\centering
\resizebox{\columnwidth}{!}{
\begin{tabular}{ccc|cc}
\multirow{2}{*}{\begin{tabular}[c]{@{}c@{}}Input\\ Representation\end{tabular}} & \multicolumn{2}{c|}{Non Far-Field}  & \multicolumn{2}{c}{Far-Field} \\ \cline{2-5} 
    & Hey-Snips  & Internal & Hey-Snips  & Internal \\ \hline
with PE    & 13.22 & 5.98 & 20.37 & 13.07   \\
w/o PE    & 17.02 & 5.74 & 21.67 & 12.40  \\
Ours    & \textbf{7.86} & \textbf{3.57} & \textbf{12.26} & \textbf{8.63}  \\ \hline
\end{tabular}
}
\end{table}

The results show that the way in which the input is presented to the model matters. The difference in performance is substantial, our representation is able to reduce the error rate by 40\% under the non far-field condition as well as Hey-Snips far-field and 34\% under the internal far-field condition compared to the patch embedding approach. The table also shows that not using patch embeddings results in worse performance on the Hey-Snips dataset but slightly better results on the internal dataset compared to using patch embeddings.

\subsubsection{Activation Function}
\label{sec:ablation_activation}
Our hyperparameter search showed that Hardswish \cite{howard2019searching} activation is more effective than GELU \cite{hendrycks2016gaussian} which is originally used in the MLPMixer model. Table \ref{tab:activation} compares the mean FRR between using Hardswish and other popular activation functions on the Hey-Snips and the internal datasets using the same procedure as section \ref{sec:ablation_input}. Hardswish performs the best overall while being simpler to implement on hardware than GELU or SiLU \cite{hendrycks2016gaussian}. 

\begin{table}[tbh]
\captionof{table}{Mean FRR (\%) at 0.3 FAs per hour for various activation functions on the Hey-Snips and Internal datasets.}
\centering
\label{tab:activation}
\resizebox{\columnwidth}{!}{
\begin{tabular}{ccc|cc}
\multirow{2}{*}{Activation} & \multicolumn{2}{c|}{Non Far-Field}  & \multicolumn{2}{c}{Far-Field} \\ \cline{2-5} 
                       & Hey-Snips  & Internal  & Hey-Snips  & Internal  \\ \hline
GELU    & 8.69 & 3.95 & 13.13 & 10.02  \\
ReLU    & 9.76 & 4.03 & 14.90 & 9.29  \\
SiLU    & 8.33 & 4.29 & \textbf{11.90} & 9.91  \\ 
Hardswish   & \textbf{7.86} & \textbf{3.57} & 12.26 & \textbf{8.63}  \\ \hline
\end{tabular}
}
\end{table}

\section{Conclusion and Future Work}
\label{sec:conclusion}

In this paper, we developed an effective small footprint QbyE-MLPMixer for open-vocabulary KWS by adapting the MLPMixer. We compared the performance of our model with other state-of-the-art RNN and CNN models to demonstrate the effectiveness of our approach. Experiments were conducted on both the publicly available Hey-Snips dataset and an internal dataset with eight keywords and 400 speakers. The results show that the MLPMixer model performs the best overall especially under noisy conditions. We also show that feeding the MFCC features directly to the model results in better performance compared to flattened patch embeddings. Finally, we show that Hardswish activation performs the best overall compared to other activation functions. To the best of our knowledge, this is the first investigation of an MLPMixer model on the QbyE open-vocabulary KWS task. The model's ability to extract information to a small embedding space can positively impact other QbyE related applications such as speaker verification and novelty detection. Future research on the rethinking of MLPs and the mixing idea could be beneficial for understanding the roles that these fundamental layers play in achieving competitive performance.


\clearpage
\bibliographystyle{IEEEtran}

\bibliography{refs}

\end{document}